# Structural, electrical and magnetic properties of nanostructured $Mn_2Ni_{1.6}Sn_{0.4}$ melt spun ribbons

Nidhi Singh[#], Barsha Borgohain, A. K. Srivastava, Ajay Dhar and H. K. Singh[*]

National Physical Laboratory, Dr. K. S. Krishnan Raod, New Delhi-110012, India

Abstract

Nanocrystalline ribbons of inverse Heusler alloy $Mn_2Ni_{1.6}Sn_{0.4}$ have been synthesised by melt spinning of the arc melted bulk precursor. The single phase ribbons crystallize into a cubic structure and exhibit very fine crystallite size of < 2 nm. Temperature dependent magnetization (M-T) measurements reveal that austenite (A)-martensite (M) phase transition begins at $T_{CA}^C \approx 248$ K and finishes at $T_M^C \approx 238$ K during cooling cycle and these values increase to $T_{CA}^W \approx 267$ K and $T_M^W \approx 259$ K while warming. In cooling cycle, the A-phase shows ferromagnetic (FM) ordering with a Curie temperature $T_{CA}^C \approx 267$ K, while both the FM-antiferromagnetic (AFM) and M-transitions occur at $T_M^C \approx 242$ K. The M-phase undergoes FM transition at $T_{CM} \approx 145$ K. These transitions are also confirmed by temperature dependent resistivity (ρ-T) measurements. The observed hysteretic behaviour of M-T and ρ-T in the temperature regime spanned by the A-M transition is a manifestation of the first order phase transition. M-T and ρ-T data also provide unambiguous evidence in favour of spin glass at $T < T_{CM}$. The scaling of the glass freezing temperature ($T_f$) with frequency, extracted from the frequency dependent AC susceptibility measurements, confirms the existence of canonical spin glass at $T < T_{CM} \approx 145$ K. The occurrence of canonical spin glass has been explained in terms of the nanostructuring modified interactions between the FM correlations in the martensitic phase and the coexisting AFM.

Keywords: Heusler Alloy, Austenite-Martensite Transition, Spin glass, Melt Spun Ribbon

[*]Corresponding author; E-mail: hks65@nplindia.org
[#]E-mail: singhnidhi@nplindia.org

**Introduction**

During the last decade full Heusler alloys, having the general formula $X_2YZ$ (X and Y are transition metals and Z is a main group element) [1-4], e.g. $Ni_2Mn_{2-x}Z_x$, $Co_2Mn_{2-x}Z_x$, etc. have gained focus because of the multiplicity of physically interesting and technologically important properties like magnetocaloric/inverse magnetocaloric effect (MCE/IMCE) [4-11], ferromagnetism (FM), and magnetic shape memory effect (MSME) [4,12-16], exchange bias (EB) [17,18], etc. The structural symmetry of these alloys is determined by the relative valance of the X and Y atomic species. When the valence of X is larger than Y ($V_X>V_Y$), the atomic sequence along the body diagonal is X-Y-X-Z and such alloys crystallize in $L2_1$ type structure consisting of four interpenetrating FCC lattices with basis vectors $\mathbf{a}(X) = (0,0,0)$, $\mathbf{b}(Y) = (1/4,1/4,1/4)$, $\mathbf{c}(X) = (1/2,1/2,1/2)$ and $\mathbf{d}(Z) = (3/4,3/4,3/4)$; prototype being $Cu_2MnAl$ [19]. However, when $V_X<V_Y$, the alloy crystallizes in the XA type cubic structure and the sequence of the atoms on the above mentioned basis vectors is changed to X-X-Y-Z. Such alloys are referred to as inverse Heusler and the prototypic structure representative of this class is $Hg_2TiCu$ [19]. The first principle electronic structure calculations of several inverse Heusler alloys have shown that the XA structure is energetically preferred over the $L2_1$ [20-23]. The structural feature that distinguishes the inverse Heusler alloys such as $Mn_2NiSn$ from the Heusler $Ni_2MnSn$ system is that the two Mn atoms occupy two crystallographically non-equivalent positions, unlike the two Ni atoms in the Heusler structure. This crystallographic non-equivalence of the Mn atoms leads to complex magnetic structures.

In Heusler alloys magnetic and structural transitions breeding multifunctionality are driven by a strong composition dependent magneto-structural coupling [11, 15, 16, 24-30]. The high temperature austenite (A) phase, which has $L2_1$ cubic structure undergoes paramagnetic to ferromagnetic (PM-FM) phase transition. At further lower temperatures the FM to anti-ferromagnetic (AFM) transition is accompanied by simultaneous first order structural phase transition from the higher symmetry A-phase to the lower symmetry martensite (M) phase. The M-phase, depending on the composition, crystallizes in either orthorhombic or monoclinic symmetry and is magnetically characterized by coexisting FM and AFM correlations. The FM interaction depends on the distance between Mn atoms occupying the Y-sublattice of the $X_2YZ$ Heusler structure [25, 26]. In Mn rich alloys, the excess Mn atoms occupy the Ni or Sn site and the strength of the AFM interaction in the M-phase depends on the hybridization between Ni and Mn 3d orbitals [26, 27]. The Ni-Mn hybridization is diluted only by the change in the Ni/Mn ratio, which in turn destabilizes the martensitic phase transition and the $L2_1$ structure. Recently Khan et al. [28] have demonstrated that the Ni-Mn hybridization is weakened by partial substitution of Mn by Cr. The weakening on Ni-Mn enhances FM interaction and lowers the martensitic transition temperature ($T_M$). Hence the dominant factor controlling the A-M phase transition is the nature of hybridization between Ni 3d and Mn 3d states of the Mn atoms at the Sn site. At T<$T_M$, the influence of the AFM interaction becomes tremendously large resulting in a magnetically frustrated state that has no magnetic long-range order and is in fact a spin glass.

Till date the main focus of research has remained on the full Heusler alloys and little experimental work has been done on their inverse variants. First principle studies have been carried out on numerous inverse Heusler alloys of the type $X_2YZ$ [$V_X<V_Y$; $X^A$(0,0,0)-$X^B$(1/4,1/4,1/4)-Y(1/2,1/2,1/2)-Z(3/4,3/4,3/4)] with X=Sc, Ti, V, Mn, Cr, etc., Y=Ni, Mn, Co, Fe, etc. and Y=Al, Si, As, Sb, Sn, etc [22,29,30,31]. When the atom X is Mn, the energy position of the d states of the $Mn^A$ and Y atoms is much closer and the nonbonding spin-down states are occupied like in the usual full-Heusler alloys, and the Slater Pauling (SP) rule for magnetization is $M_t = Z_t - 24$ [22,30]. However, a modified SP rule, $M_t = Z_t - 28$, has also been suggested for these alloys [22]. The predicted lattice constant of the XA type cubic structure of the Mn based inverse Heusler alloys is a≈5.9 Å for Al and is expected to decrease (increase) with decreasing (increasing) atomic radii of the Z element [22]. The nature of spin coupling between the nearest neighbours $Mn^A - Mn^B$ is still not clear and appears to depend on the nature of the Z element [29]. In fact, Huo et al. [29] reported a weakened AFM ordering of $Mn^A - Mn^B$ moments in $Mn_2NiSn$. In usual Heusler alloys this ordering tends to be AFM type when they are nearest neighbours, while in inverse systems AFM is weakened through a decreases in the lattice constants. In $Mn_2NiSb$ the moment of the two Mn atoms couple ferromagnetically, while $Mn_2NiGa$ appears to have AFM coupling [30]. For $Mn_2NiSn$ system several interesting features such as, experimentally measured $T_C$=565 K and saturation moment ($M_S$) was found to be 2.95 $\mu_b$/fu at (5 K), which is close to the theoretical value of 3.35 $\mu_b$/fu [29,30]. The experimental investigations of $Mn_2YZ$ (Y=Co, Ni Y=Al, Ga, In, Ge, Sn, Sb) and related its first principle calculations carried out by Liu et al.[31,32] appear to have been successful in clarifying several structural issues. They confirmed that in $Mn_2NiZ$ and $Mn_2CoZ$, Y (Ni/Co) atoms preferentially occupy $X^A$ (0,0,0) and Y (1/2,1/2,1/2) sites and crystallize in XA type $Hg_2CuTi$ cubic structure (SG F$\bar{4}$3m) and that two Mn atoms by virtue of being the nearest neighbours provide a fair degree of complexity to the magnetic interactions. The above findings suggest that the fraction of Ni atoms could occupy the Mn (0,0,0) site and give rise to Ni-Mn hybridization and as mentioned above would favour the AFM ordering in the M- phase.

The complexity of the structural and magnetic ordering of the transition metals in the Mn$_2$YZ having XA type Hg$_2$CuTi cubic structure suggests that even a small disorder either intrinsic or extrinsic can drastically modify their magnetic interactions. The consequences of extrinsic disorder on the structure and magnetic properties of the inverse Heusler alloys have not been paid much attention [7, 33, and 34]. One way to introduce extrinsic disorder is to enhance surface effects by nanostructuring. However, to be sure that observed modifications in magnetostructural properties are mainly due to the extrinsic disorder, the intrinsic atomic order should remain unmodified. Spin melting of arc melted precursor material [32] is one possible way to achieve nanostructuring without appreciably modifying the intrinsic atomic order.

In the present work, the consequences of nanostructuring on the magnetic transitions have been studied in melt spun ribbons of Mn$_2$Ni$_{1.6}$Sn$_{0.4}$ inverse Heusler alloy. It may be mentioned here that there are very few studies on the structure and magnetotransport properties of melt spun ribbons of Mn rich inverse Heusler alloys [7]. Our study demonstrates the appearance of A-FM to M-AFM and M-AFM to M-FM phase transition and spin glass (SG) in the M-phase. It was observed that in contrast to the bulk alloy of similar composition, the first order nature of the A-FM to M-AFM phase transition is more pronounced in these ribbons [34]. The enhanced nature of the first order phase transition and origin of the SG behaviour in these alloys could be traced to nanostructuring enhanced competition between the coexisting FM and AFM interactions. The frequency dependent ac susceptibility and frequency dependence of the characteristic temperature, in the frame work of the Vogel-Fulcher and the dynamic scaling law, confirmed the canonical nature of the SG in these ribbons.

**Experimental**

Mn (Alpha Aesar-99.99%), Ni (Alpha Aesar-99.99%) and Sn (Alpha Aesar-99.99%) granules were taken in a proper chemical ratio so as to yield a composition of Mn$_2$Ni$_{1.6}$Sn$_{0.4}$. The material was made employing arc melting furnace (Edmund Bhuler, Germany) and the melting was repeated several times in order to achieve homogeneity. The Mn$_2$Ni$_{1.6}$Sn$_{0.4}$ alloy thus obtained was melt-spun (Edmund Bhuler, Germany) to obtain ribbons of thickness 25-30 μm. Structural characterization was performed on the melt spun alloy by powder X-ray diffraction (XRD, model Rigaku-miniflex, Cu-K$_{\alpha 1}$, 1.5406 Å) and the high resolution transmission electron microscopy (HRTEM, model: JEOL JEM 2100 F, with field emission electron gun assisted and operated at 200 keV). Temperature dependent resistivity was measured in close cycle refrigerator attached to a Keithley delta mode measurements system and Lakeshore cryogenic temperature controller. The sample was cooled/warmed at 2 K/minute. The magnetic measurements like zero filed cooled (ZFC), field cooled cool (FCC) and field cooled warming (FCW) dc magnetization were measured by a commercial SQUID magnetometer (Quantum Design).

**Results and Discussion**

The XRD pattern of $Mn_2Ni_{1.6}Sn_{0.4}$, presented in Fig. 1, clearly shows the occurrence of three distinct types of diffraction maxima [31]. The first is type I superlattice peaks (h, k, l all odd) represented by (111) and (311) maxima. The second is type II superlattice reflection (200) for which h+k+l=2n and h, k, l, all even with n=1, 2, 3… In addition to these superlattice maxima, type III principal reflections like (220), (400), etc. for which h+k+l= 4n and h, k, l, all even with n=1, 2, 3... exists. The type I superlattice reflections are representative of order between $X^B$(1/4,1/4,1/4) and Z(3/4,3/4,3/4) sublattices, while the type II superlattice reflections arise from the ordering of the $X^A$(0,0,0), Y(1/2,1/2,1/2) and $X^B$(1/4,1/4,1/4) sublattices. In view of this, the higher intensity of the (111) peak with respect to the type II (200) peak could be taken as the representative of a higher degree of Mn-Sn ordering (than Ni-Mn ordering). The lattice parameter of the cubic unit cell has been estimated to be $a_c$=0.5978 nm and is close to the theoretically predicted value [22]. The above diffraction characteristics have also been reported in $L2_1$ type (prototype $Cu_2MnAl$) as well as XA type ($Hg_2CuTi$) cubic structures [2, 3].

The local area structural and microstructural features of melt spun $Mn_2Ni_{1.6}Sn_{0.4}$ ribbon were also investigated by HRTEM and are shown in Fig.2. It was observed that in general the solidified microstructure was mushy with an ultra-fine grains abutting to their adjacent grains throughout the sample (Fig. 2a). However, at high magnifications the microstructure observed shows individual grains and are consisted of clear boundaries and facets, (Fig. 2b). The observed grain-size distribution was very narrow with grains ranging between 1 to 1.5 nm (inset in Fig. 1a). At an atomic scale, the grains were found to be randomly oriented with respect to one another. In the inset of fig. 1b, two sets of planes with interplaner spacings of 0.25 and 0.21 nm corresponding to hkl values of 211 and 220, are observed. The different orientation of crystallographic planes was observed over the entire specimen. As another illustrative example, Fig. 2c shows two sets of planes with interplaner spacings of 0.35 and 0.21 nm corresponding to hkl values of 111 and 220. A magnetic characteristic of the material influences the atomic scale images and instead of sharp planes, the magnetic anisotropy results into thick fringes, as elucidated in the inset of Fig. 2c. Fig. 2d displays a selected area electron diffraction pattern (SAEDP) along [$\bar{1}$13] zone axis of cubic (bcc) crystal and the corresponding diffraction-spots of the planes, 220, 301, 2$\bar{1}$1 and 1$\bar{2}$1 are indexed on the pattern (Fig. 2d). The inset in Fig. 2d, further elucidates a typical SAEDP recorded along [$\bar{1}$11] zone axis of bcc crystal structure with important planes of hkl: 01$\bar{1}$, 110 and 101, in reciprocal space. In addition to the ultra-fine grains, it was noted that the electron diffraction pattern appeared predominantly along the [$\bar{1}$13] zone of bcc crystal structure, which is presumably due to a preferred-oriented growth of the grains during rapid solidification employing melt spinning.

The temperature dependence of the resistivity ($\rho$ - T) at zero magnetic field (H=0) measured in cooling and warming cycles, is plotted in Fig. 3. The high temperature A-phase shows a metal like $\rho$ - T behaviour and as the temperature is lowered below the room temperature, it transforms to the M-phase, which is marked by a rise in the $\rho$ - T curve (Fig. 3). As the temperature is lowered further the resistivity appears to saturate with temperature. It is evident from this figure that the A-M transition begins at $T_{CA}^{C} \approx 248$ K and culminates at $T_{M}^{C} \approx 238$ K. The transition temperatures, determined from the d$\rho$/dT-T data, are shown in the inset of Fig. 3. In the warming cycle these transitions shift to appreciably higher temperatures, e.g., the M-transition appears at $T_{M}^{W} \approx 259$ K and the M-A transition finishes at $T_{CA}^{W} \approx 267$ K. The cooling-warming $\rho$ - T curves (Fig. 3) show a reversible behaviour in the lower temperature regime (T<50 K) and above that it becomes irreversible. Further this irreversibility increases with temperature and maximizes in the vicinity of M-A transition. This irreversible behaviour clearly shows the first order nature of the M-A phase transition.

The temperature dependence of DC magnetization of the Mn$_2$Ni$_{1.6}$Sn$_{0.4}$ ribbon, measured employing ZFC, FCC and FCW protocols reveals several interesting features (Fig. 4). All the transition temperatures have been determined from the first order temperature derivative of magnetization $\left(\frac{dM}{dT}\right)$ and hence correspond to the midpoint of the respective transitions. The $\frac{dM}{dT} - T$ data is plotted in the inset of Fig. 4. The rise in the M (T) just below the room temperature signals the FM ordering in the austenite phase with a Curie temperature $T_{CA}^{C} \approx 267$ K. The sharp decline of the M (T) at T<261K marks the appearance of the AFM correlations concomitant with the A-M transition at $T_{M}^{C} \approx 242$ K. As T is lowered further, the M (T) curve changes slope and rises due to the appearance of FM order in the M-phase, characterized by a Curie temperature $T_{CM} \approx 145$ K. As T is lowered below $T_{CM}$ the ZFC-FCW M (T) start diverging with the latter declining quite sharply at T$\approx$100 K. Such divergence is a consequence of the reappearance of AFM correlations in the M-phase and the sharp decrease in M (T) is a signature of spin freezing. These features have been recognized as standard attributes of a glass like state [35, 36]. The FCW remains reversible with respect to the FCC M (T) up to T≈160 K and above this a prominent irreversibility is observed. The martensitic transition and the Curie temperature of the austenite phase are shifted to $T_M^W$ ≈257 K and $T_{CA}^W$≈268 K respectively in the warming cycle. The well-developed FCC-FCW hysteresis in the temperature range 160 K - $T_{CA}^W$ is a signature of the first order coupled magnetostructural transition from room temperature FM-A to lower temperature AFM-M phase. The reversible behaviour of the FCC-FCW curves in the lower temperature regime could be attributed to the presence of a magneto-structural glass. It is interesting to note that the value of the austenite Curie temperature $T_{CA}$ of our ribbon samples is nearly same as reported by Ma et al.[34] for bulk sample of identical composition. However, the noticeable difference between their bulk and our ribbon samples is the pronounced FCC-FCW hysteresis, which could be due to the enhanced AFM fraction. It has been suggested earlier that a stronger Ni-Mn hybridization favours a stronger AFM exchange [26, 28, and 37]. The AFM interactions, as they are controlled by Ni-Mn hybridization, may not vanish completely and hence their presence in the FM matrix could give rise to a glassy behaviour.

In Mn-rich Heusler alloys, the change in M (T) just below the A-M phase transition is correlated to the AFM interaction active in the M-component of the two phase mixture. Once the complete M-phase is achieved, the AFM correlations diminish and are in fact substituted by FM interaction, as is clear by the rise in magnetization below T≈160 K (Fig.4). The reappearance of AFM correlations in the martensitic phase could be attributed to the coupling between the excess Mn atoms [29-31]. The AFM coupling between the two Mn sublattice atoms is further strengthened by the reduced Mn-Mn distance due to nanostructuring. In the present case, the ultrafine grain structure in $Mn_2Ni_{1.6}Sn_{0.4}$ ribbon clearly suggests an enhanced contribution from the surface disorder. Such interfacial disorder could enhance magnetostructural frustration that in turn could strengthen the AFM interaction. This type of AFM order that appears under a FM background and the associated glass transition as evidenced in the present case by the reversible (irreversible) behaviour of the FCC-FCW (ZFC-FCW) M (T), is regarded as re-entrant [35].

The nature of the magnetic phase in the present melt spun $Mn_2Ni_{1.6}Sn_{0.4}$ ribbons was further investigated by ac susceptibility measurements. The ac susceptibility was measured at several frequencies and field amplitude of $H_{ac}$=10 Oe in absence of a dc magnetic field ($H_{dc}$=0). The measured real (M′) and imaginary (M″) components of the ac susceptibility are plotted in Fig. 5 and Fig. 6, respectively. M′ - T curve shows two clear peaks, the higher temperature peak (≈253 K), which does not exhibit any frequency dispersion, characterizes the Curie temperature of the A-phase. The peak in the temperature dependent ac susceptibility, showing clear frequency dispersion, is referred to as the freezing temperature, $T_f$ and its experimentally measured value is slightly lower for the imaginary component of the ac susceptibility curve. $T_f$ is generally close to the ZFC-FCW divergence in the dc M (T) data. As seen in Fig. 5, the lower temperature (≈130 K) peak, observed in the vicinity of the ZFC-FCW divergence, shifts towards higher temperatures with a concomitant decrease in the peak amplitude. The value of $T_f$ is found to be 130.74 K at 33 Hz, which increases to $T_f$ = 135.21 K at 9999 Hz. M″ - T curve shows a peak corresponding to $T_{CA}$ at T≈250 K, which is slightly lower than that of the M′ - T curve. The frequency dispersion of the lower temperature peak in M″ - T curve reveals a shift in $T_f$ ($T_f$ = 119.68 K at 33 Hz) towards higher temperature with increase in the frequency with a simultaneous increase in the peak amplitude (Fig. 6). The observed frequency dispersion in $T_f$ in the ac susceptibility curves in conjunction with the prominent divergence in the ZFC-FCW M (T) curve clearly demonstrates the existence of a spin glass state in the $Mn_2Ni_{1.6}Sn_{0.4}$ ribbons. Spin glass is generally regarded as a consequence of frustration caused by competing magnetic interactions and competing geometries [35, 36]. In cases when the grain size is very fine, even particulates of FM domains may also exhibit a spin glass like behaviour. The sharp drop in M″ (T) of the ac susceptibility at T<$T_f$ (much sharper than that in the M′ (T) curve) is a signature of the magnetic system with blocked magnetic clusters. This also suggests that such a glassy transition involves groups of coherent spins rather than individual atomic spins. It is also interesting to note that the M″ - T curve shows a small peak around T≈38 K features at higher frequencies. This peak shifts towards higher temperatures and grows in amplitude as frequency is increased. This could be attributed to the local fluctuations in the spin glass. In view of our dc M (T) data and the ac susceptibility results presented above the origin of the spin glass in $Mn_2Ni_{1.6}Sn_{0.4}$ ribbons could be traced to a combined consequence of the competing magnetic (FM/AFM) interactions and the geometrical frustration created by the ultrafine particulates, which could further enhance the magnetic frustration in the lower temperature M-phase.

In order to gain more information about the nature of the magnetic state that appears to be a spin glass resulting from the combined effect of magnetic and geometric frustration we calculated the dimensionless quantity (Φ) given by

$$\Phi = \frac{\Delta T_f}{T_f \Delta \log f}$$

For typical spin glass systems 0.005<Φ<0.08, while super paramagnetic systems are characterized by Φ≈0.1 [35, 36]. Our experimental data yields Φ≈0.014, which is typical of canonical SG alloys, like, PdMn, NiMn and AuFe and hence is perfectly in the limits to qualify the low temperature magnetic state of $Mn_2Ni_{1.6}Sn_{0.4}$ ribbons as a spin glass [36].

The frequency dispersion of $T_f$ provides important information about the nature of the magnetic clusters and their interaction in the spin glass state. The empirical Vogel-Fulcher law originally proposed to describe the behaviour of supercooled liquids, [35, 36, 38, 39] has been successfully used in our samples to decipher the nature of the spin glass state. The Vogel-Fulcher law, as applied to spin glass systems, is expressed as

$$\tau = \tau_0 e^{\frac{E_a}{k_b(T_f - T_0)}}$$

Here, $\tau=1/f$, $E_a$ $k_b$ is Boltzmann constant is the activation energy associated with the spin glass clusters and $T_0$ is the Vogel-Fulcher parameter related to the interaction between the magnetic clusters. The observed frequency dispersion behaviour in the V-F law is presented on a $\ln(\tau)$ – $1/(T_f-T_0)$ plot in Fig. 6 for both the real and imaginary components. Both the real and imaginary part of the data are well traced by a linear fit. The real part of the data yields $\tau_0≈1.3\times10^{-9}$ sec, $T_0 ≈ 122$ K and $E_a ≈13.3$ meV. The imaginary data yields slightly lower value of $T_0≈109$ K but slightly higher values for $\tau_0≈8.3\times10^{-9}$ sec and $E_a≈13.8$ meV. The value of $E_a$, obtained in the present case, is considerably larger than that reported by Ma et al. for bulk $Mn_2Ni_{1.6}Sn_{0.4}$ bulk alloy [34]. This could be interpreted as a manifestation of much stronger inter-cluster interaction in the ribbon form of $Mn_2Ni_{1.6}Sn_{0.4}$ than in the corresponding bulk. The inter-cluster interaction could be enhanced by the additional geometric frustration caused by the reduced particulate size in the melt spun ribbons. The value of $\tau_0$ obtained from the Vogel-Fulcher law in the present study for the $Mn_2Ni_{1.6}Sn_{0.4}$ ribbons is considerably smaller than that reported by Ma et al. [34] and Chatterjee et al. [33] for $Mn_2Ni_{1.6}Sn_{0.4}$ and $Mn_2Ni_{1.36}Sn_{0.64}$ bulk variants but nearly of the same order as that for the prototype SG system like CuMn (4.6 at %) alloy [36]. This shows that the SG state in nanostructured ribbons is much closer to the canonical SG systems.

The spin glass systems have also been treated in the framework of the standard theory of dynamic scaling in the vicinity of phase transition [35, 36]. In this frame work the critical relaxation time follows a power law divergence

$$\tau = \tau_0 \left(\frac{T_f - T_g}{T_g}\right)^{-z\nu}$$

Here, ν is critical exponent of the correlation length and z is the dynamic exponent. The parameter $T_g$ is taken as the static glass transition temperature. The $T_f$ dispersion as observed in the real and imaginary components of ac susceptibility was analysed in terms of the dynamical scaling law mentioned above. This is brought about by the ln(τ) – [($T_f$-$T_g$)/$T_g$] data plotted in Fig. 6. Both the set of data are well traceable by a linear fit and the three fitting parameters were evaluated from the best fits to the observed data. For the real part $T_g$ = 127 K, $\tau_0 \approx 7\times10^{-13}$ sec and zν=6.95. The imaginary component of the yields $T_g$ = 115.7 K, $\tau_0 \approx 6\times10^{-11}$ sec and zν=5.9. These values are well within the range of typical spin glass systems, e.g., 4<zν<12 typically. For CuMn (4.6 at. %), $\tau_0 \approx 10^{-12}$ sec and zν=5.5 with $T_g$=27.5 K [36]. This further confirms the canonical SG behaviour in the present spin-melted ribbons.

**Conclusions**

Nanocrystalline ribbons of inverse Heusler alloy $Mn_2Ni_{1.6}Sn_{0.4}$ have been synthesised employing melt spinning of arc melted alloy. These ribbons, which exhibit a very fine grain size are homogeneous and structurally ordered. The temperature dependent resistivity and magnetization, both show unambiguous signature of a first order (hysteretic) austenite-martensite phase transition. In the martensitic phase a reentrant FM phase evolves in the presence of the AFM phase. The AFM phase appears to be modulated by the enhanced geometrical frustration due to nanostructuring and gives rise to a spin glass. The scaling of the freezing temperature with frequency in the frame work of the Vogel-Fulcher law, the dynamic scaling law and the observed values of the time constants, clearly show that in sharp contrast to their bulk counterpart the spin glass state in the $Mn_2Ni_{1.6}Sn_{0.4}$ ribbons is canonical in nature.

**Acknowledgements**

Constant support and encouragement from Prof. R. C. Budhani, Director, CSIR-NPL, India is gratefully acknowledged. Authors would like to thank CSIR for financial support. One of the authors BB would like to thank DST-INSPIRE for financial support. Authors are grateful to Dr. Anurag Gupta and Dr. V. P. S. Awana for magnetic measurements and for technical support to Radhey Shyam and N K Upadhayay.

**Figure Captions**

Figure 1:  XRD pattern of $Mn_2Ni_{1.6}Sn_{0.4}$ of melt spun ribbon acquired at room temperature employing Cu-K$\alpha$ radiation.

Figure 2:  HRTEM micrographs of $Mn_2Ni_{1.6}Sn_{0.4}$ melt spun ribbon showing (a,b) fine grained microstructure of the melt spun alloy,(c) atomic scale images and (d) SAEDP along [$\bar{1}13$] zone axis of bcc crystal structure. Insets: (a) grain size distribution, (b) atomic scale image, (c) thick fringes due to magnetic characteristic, and (d) SAEDP along [$\bar{1}11$] zone axis of bcc crystal.

Figure 3:  Temperature dependent resistivity of $Mn_2Ni_{1.6}Sn_{0.4}$ ribbons measured in the cooling and warming cycles at zero magnetic field. The inset shows the temperature derivate of resistivity used to determine various transition temperatures.

Figure 4:  Temperature dependent magnetization of $Mn_2Ni_{1.6}Sn_{0.4}$ ribbon measured employing ZFC, FCC and FCW protocols at applied dc field of H=100 Oe. The inset shows the temperature derivate of FCC and FCW magnetization used to determine various transition temperatures.

Figure 5:  Temperature dependence of the real part of the ac susceptibility measured at different frequencies. The inset is a magnified image of the region around the lower peak and shows the frequency dispersion of $T_f$. The arrow marks the direction of increasing frequency.

Figure 6:  Temperature dependence of the imaginary (dissipative) part of the ac susceptibility measured at different frequencies. The inset is a magnified image to highlight the frequency dispersion of $T_f$. The arrow marks the direction of increasing frequency.

Figure 7:  The agreement between the observed frequency dispersion of the freezing temperature $T_f$ and the Vogel-Fulcher law on $\ln(\tau)$-$(T_f-T_0)^{-1}$ plot. The black and spheres are $T_f$ obtained from the real and imaginary part of the ac susceptibility. The solid line are best fit to the experimental data.

Figure 8:  The agreement between the observed frequency dispersion of $T_f$ and the dynamic scaling law on $\ln(\tau)$-$(T_f-T_g)/T_g$ plot. The black and blue spheres are $T_f$ obtained from the real and imaginary part of the ac susceptibility.

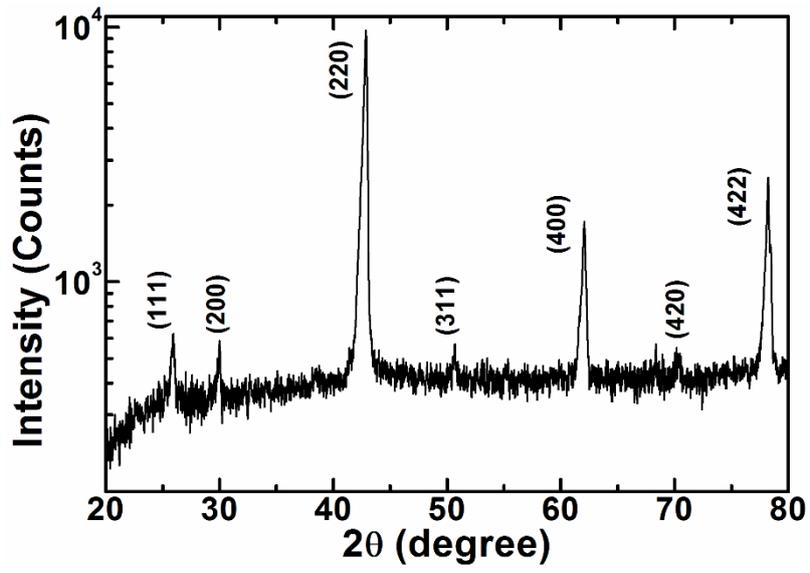

Fig. 1

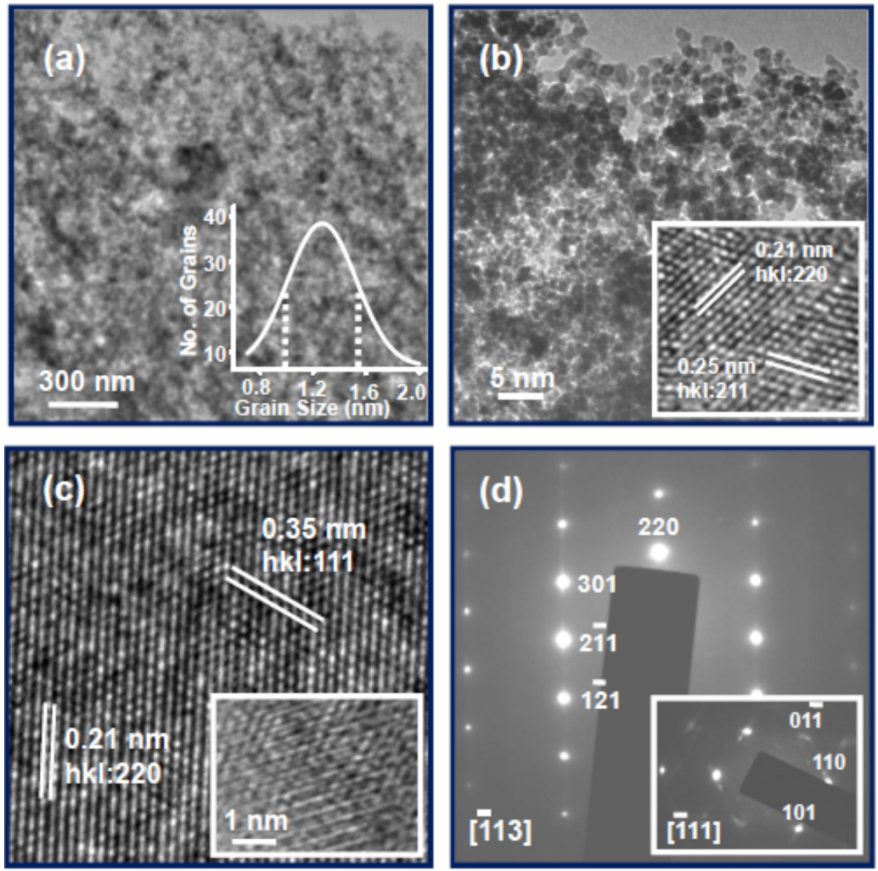

Fig. 2

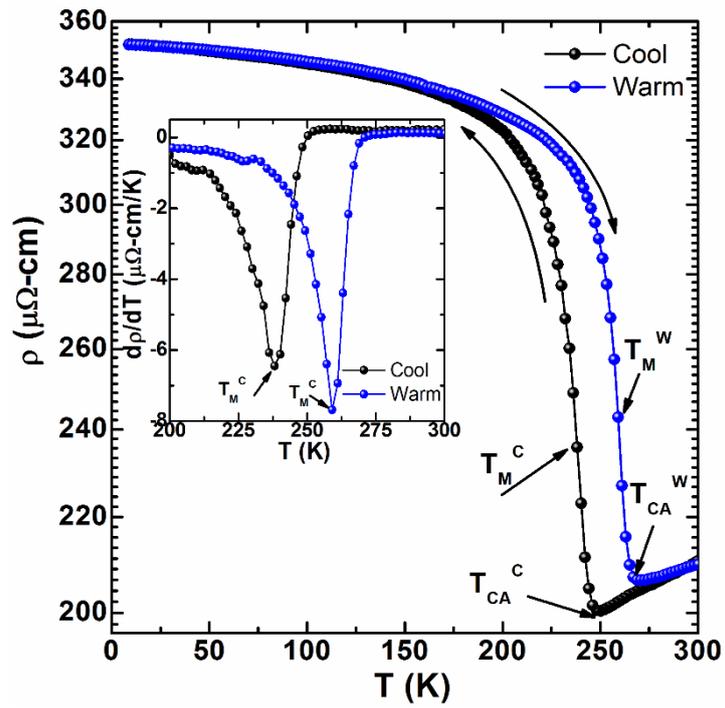

Fig. 3

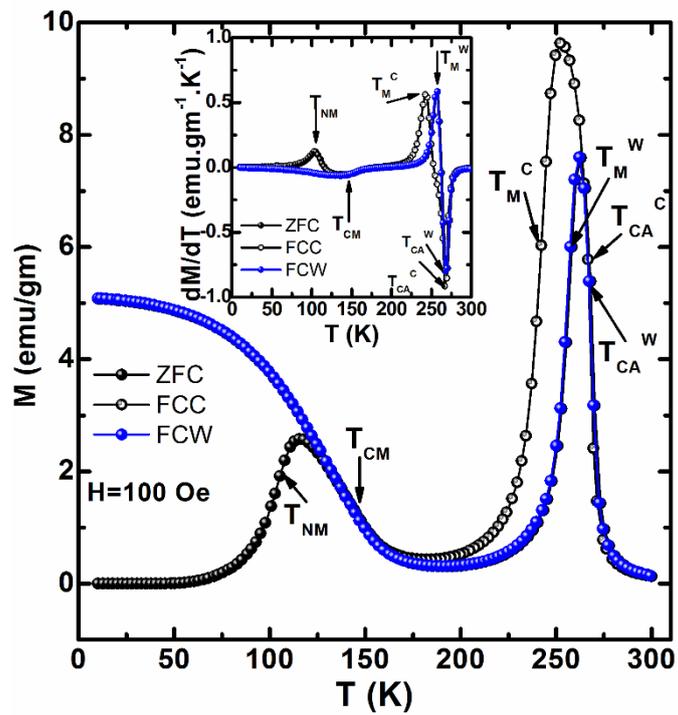

Fig. 4

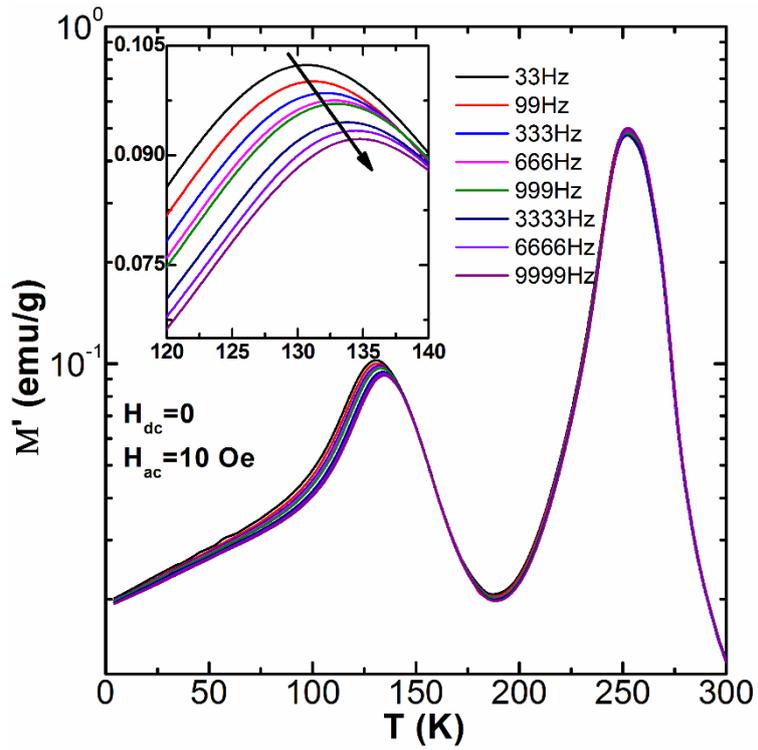

Fig. 5

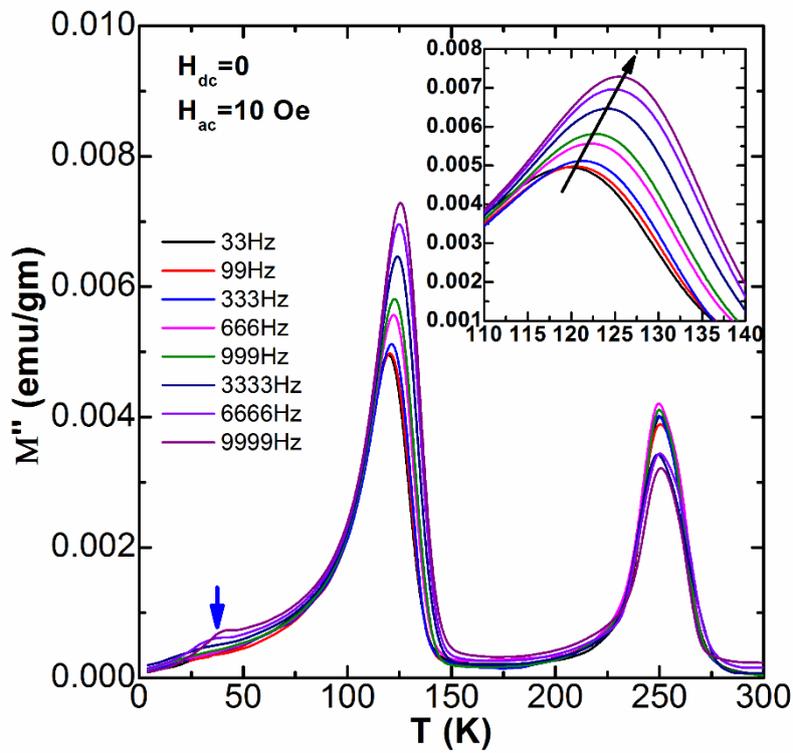

Fig. 6

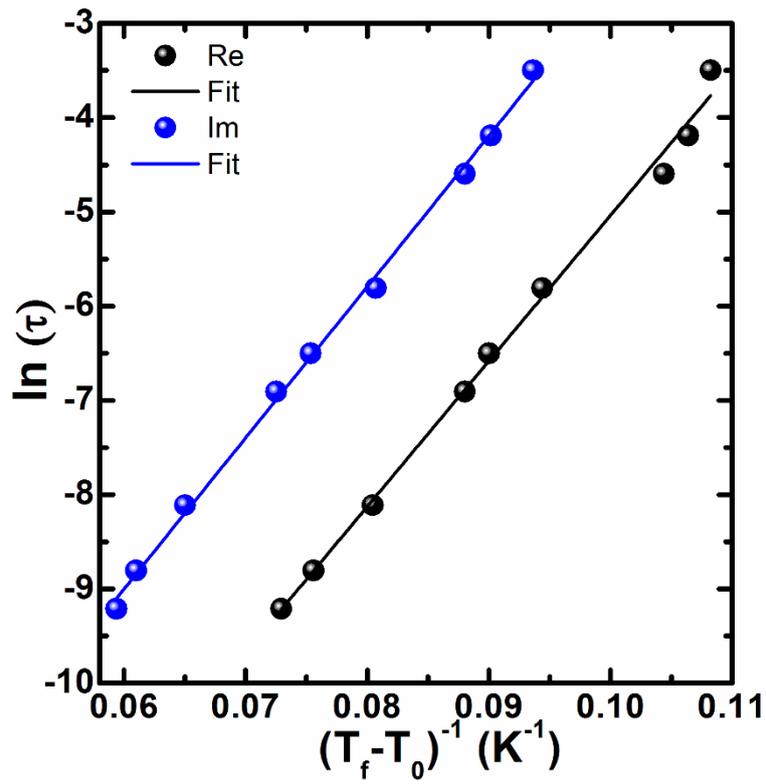

Fig. 7

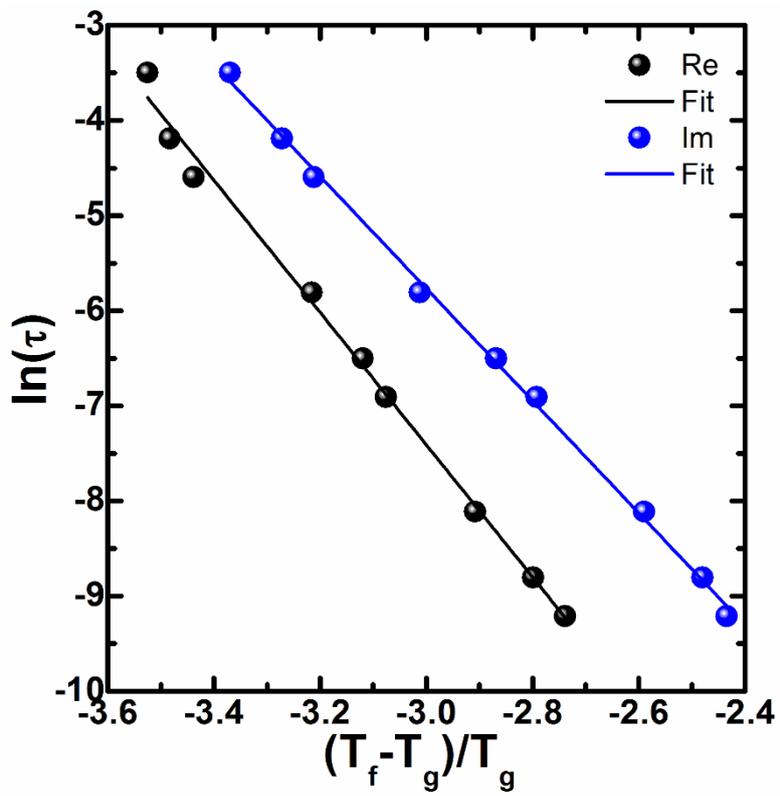

Fig. 8